THE INTERACTION OF REGIONAL AND LOCAL IN THE DYNAMICS OF THE COFFEE
RUST DISEASE

John Vandermeer[1]

Pejman Rohani[1,2]

1. Department of Ecology and Evolutionary Biology

University of Michigan

Ann Arbor, MI 48109

2. Center for the Study of Complex Systems

University of Michigan

Ann Arbor,  MI 48109

**Abstract**

A simple mean field model is proposed that captures the two scales of dynamic processes involved in the spread of the coffee rust disease. At a local level the state variable is the fraction of coffee plants on an average farm that are infested with the disease and at the regional level the state variable is the fraction of farms that are infested with the disease. These two levels are interactive with one another in obvious ways, producing qualitatively interesting results. Specifically there are three distinct outcomes: disease disappears, disease is potentially persistent, and disease is inevitably persistent.  The basic structure of the model generates the potential for "tipping point" behavior due to the presence of a blue sky bifurcation, suggesting that the emergence of epizootics of the disease may be unpredictable.

The coffee rust disease, caused by the fungus *Hameilia vastatrix*, emerged as a major problem in the late nineteenth century where it forced the abandonment of coffee production in large areas of southern Asia (Pendergast, 1999; McCook, 2006). Its appearance in the new world began in the 1980s (Fulton, 1984) where it rapidly expanded to all coffee producing areas but never reached the devastating levels it had earlier in southern Asia (Vandermeer et al., 2009).  This relatively benign regime seems to have changed in the 2012/2013 season when throughout the northern part of its range in Latin America, coffee farms were devastated by the disease (Cressey, 2013).  Responses to this dramatic and sudden epizootic were based on current knowledge, which is arguably incomplete from an ecological point of view.  Many policy makers focus on only a small scale when trying to understand and react to the problem, searching for advice to give to local farmers. Yet there is an evident need to also consider regional factors since the disease is spread broadly by wind currents (Kushalappa and Eskes, 1989; Schieber, 1975). Indeed, the dynamics of this disease are well known to have important components at two distinct scales, one at the level of individual coffee bushes and the other at country or even continental levels (Kushalappa and Eskes, 1989; Avelino et al, 2004; 2006; 2012).  Here we develop a toy model that incorporates both of these scales in a form that is intuitive in formulation and provides heuristic ideas as to how various management options might either interfere with one another or potentially act synergistically to deal with the problem. Although we offer no particular instantiation of the model, its qualitative behavior should be of interest to anyone concerned with managing this disease.

The basic biology of the coffee rust is well-known (Kushalappa and Eskes, 1989; Avelino et al., 2004; Vandermeer et al. 2009), although its connections, both direct and indirect, with surrounding ecosystem elements remain obscure (Vandermeer et al., 2010; Jackson et al, 2009; 2012; Avelino et al., 2012). Uredospores germinate within a drop of water on the underside of the leaf and penetrate the leaf through stomata, growing extensively in intercellular space. Penetration of cells follows with the formation of haustoria, the absorptive organ, which draws nutrients from the cell. Production of fruiting bodies follows with uridia on the surface, forming the yellow rust texture. Uredospores are then dispersed partly by touch and splash to neighboring leaves and eventually taken up by the wind to altitudes up to 1000 meters and carried at least 150 km from the initial infection (Schieber, 1972). The disease organism is potentially attacked by at least 10 different species of mycoparasities (Vandermeer et al., 2009; Carrión and Rico-Gray, 2002; Arriola et al., 1998; Jackson et al., 2012). The essential ecological features for purposes of developing the model is the long distance dispersal by wind, the local dispersal by touch and splash, the mycoparasities and other potential antagonists, and the need for a droplet of water for germination. The essential sociopolitical features include economic and political forces that cause coffee farming to either be undertaken or abandoned in a whole region.

The basic model considers two simple variables, $x$ = the proportion of coffee bushes on a particular farm that are infected with the disease and $y$ = the proportion of farms that are infected in the entire region. We presume that spore dispersal is extensive in the area (Schieber, 1972) and thus occurs at the local level as propagule rain. We further presume that the generalized infection rate at the regional level is proportional to the average intensity of infection at the local level. With these assumptions the simple model is;

$$\frac{dx}{dt} = ay(1-x) + mx(1-x) - ex$$

$$\frac{dy}{dt} = M\left(nx\right)y(1-y) - E(p+nx)y$$

where *a* is the rate of propagule rain, *m* is the infection rate from coffee bush to coffee bush and *e* is the recovery rate at the bush level, *M* is the infection rate at the regional level (the rate at which new farms become infected), *n* is a scaling factor to convert the average infection per farm into an overall availability of spores, *E* is the extinction rate of whole farms where *Ep* is the extinction rate independent of the disease itself (i.e., due to sociopolitical or local ecological factors) and *Enx* is the extinction rate due to the intensity of the disease in the region. As defined above, *x* represents the infection of bushes on an average farm (average proportion of bushes infected), whereas y represents the infection of farms in the whole region (proportion of farms infected).

Equations 1 have simple dynamics. If we presume that *p* is zero, which is to say, there is no farm failure that is independent of the disease, the basic dynamics are illustrated in figure 1. The equilibrium value of *y* is,

$$y^* = \frac{M-E}{M}$$

and, at lower values of *x*, the equilibrium value of *x* is approximately,

$$x^* = \frac{ay^*}{m-e+y^*} ,$$

whence, by examination we see that the intensity of the disease at a local level (*x*) will increase as *a* and *e* increase, but decrease as *m* increases. These results are intuitively obvious.

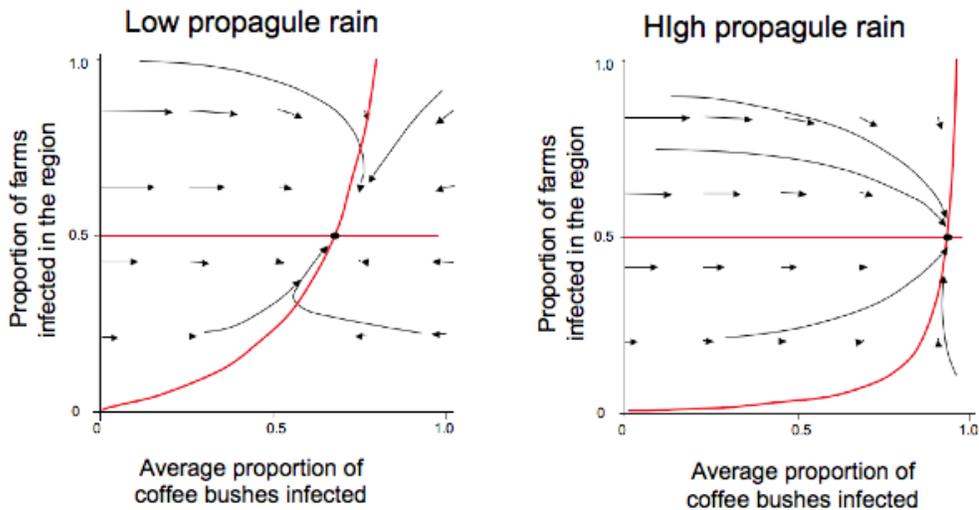

*Figure 1. Behavior of the model (equation set 1) under two parameter sets. Bold red curves are the zero net growth isoclines, smooth curves are four distinct trajectories and the collection of small arrows indicates the vector field. a. parameters set to simulate relative control of the disease (i.e., less than 100% of the bushes infected). Parameter values are a = .15; m = .01; e = 0.05, M = 1; n = 2; E = .5 and p = 0.  b. parameters set to simulate a strong epizootic (>90% of the bushes infected). Parameter values the same as in a) except a=3.*

The situation changes significantly if *p* >0, which is to say, if there is a reason other than the rust disease for a farm to either be abandoned or change production to some other crop.  For example, it is reasonable to expect farmers to abandon coffee cultivation when market prices are declining, or if some other pest in the system becomes important, or if political events create insecurity.  Any such forces are formally independent of the intensity of coffee rust and thus are modeled by *p*> 0. In figure 2, we illustrate the three qualitatively distinct results for this situation.

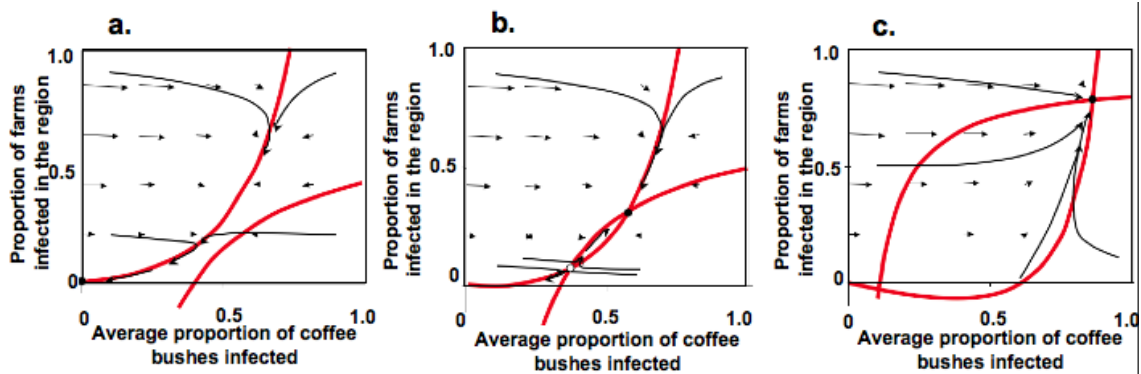

*Figure 2. Behavior of the model (equation set 1) under alternate parameter sets. Bold red curves are the zero net growth isoclines, smooth curves are four (or six) distinct trajectories and the collection of small arrows indicates the vector field. Solid black circle indicates stable node, open circle indicates saddle point. Parameters here chosen to illustrate alternative equilibria and the origin (or decay) of the disease. a) the disease remains under control no matter where it is initiated. Parameter values are a = 1, m = 0.35; e = 0.45, M = 0.35; n =1; E = 0.1 and p = 1. b) Subsequent to a blue sky bifurcation in which two equilibrium points are created (a saddle and a stable node), in which the disease can be maintained under control as long as it starts at a low mode, but there is a tipping point above which the disease becomes epizootic. Parameter values are a = 1, m = 0.6; e = 0.5, M = 0.4; n =1; E = 0.1 and p = 1. c) The situation in which the disease will be epidemic under all initial conditions. Parameter values are a = 1, m = 0.5; e = 0.2, M = 1; n =1; E = 0.1 and p = 1.*

The move from complete and automatic control of the disease (figure 2a) to the potential for control (figure 2b) is, formally, a point of bifurcation, in which alternative equilibrium points emerge "out of the blue" (and thus the appellation blue sky bifurcation). Although formally tractable (it is possible to solve for the situation of precisely two roots, one of which is negative for *x* and thus of no interest, the other of which represents the precise point of the saddle/node bifurcation), the solution is complicated and provides little insight. On the other hand, the sudden change from the possibility of control (figure 2b) to no possibility of control is evidently the point at which,

$$\frac{p}{n}\left(\frac{E}{M-E}\right) < 1 - \frac{e}{m}.$$

In sum, there is a qualitative generalization that driving parameters enable a change from 1) no possibility for the disease to become epidemic (figure 2a), to a tipping point at which the disease is either controlled or not, depending on initial conditions (figure 2b), to another type of tipping point in which the disease cannot be controlled at all (figure 2c). A simple bifurcation diagram with *p* as the tuning parameter (any of the other parameters could have also been used) is of considerable interest, as illustrated in figure 3.

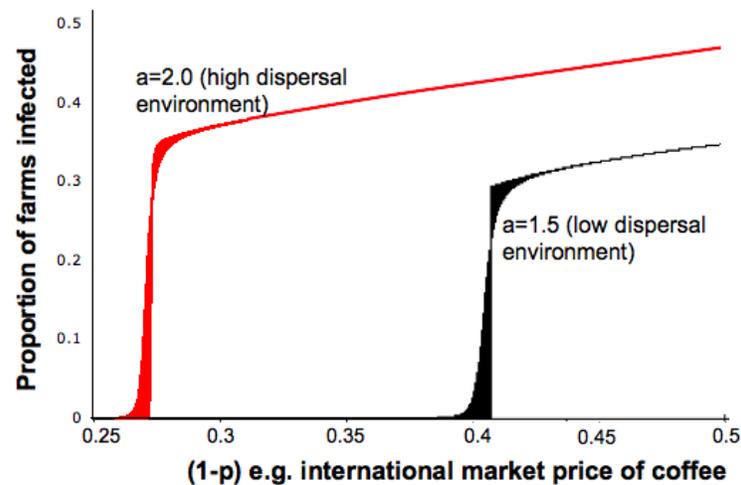

*Figure 3. Proportion of farms infected as a function of a socioeconomic driver (the example is international market price). As market price increases, p decreases (i.e., as prices go up, fewer farms go bankrupt due to economic forces), but as the number of farms increases, the probability of rust dispersion also increases. The result is a tipping point at which the region as a whole switches from relatively free of the disease, to widespread infection. The critical value of the socioeconomic driver depends on the dispersal environment of the disease (e.g., a large proportion of sun coffee in the region will result in a higher value of a and a tipping point at a lower value of the socioeconomic driver). At each value of 1-p the first 50 transients were discarded and the next 100 plotted.*

The result is that a distinct tipping point emerges, as would be expected from a qualitative glance at the zero net growth isoclines of the system (i.e., figure 2). Furthermore, the position of that tipping point depends on the background ecological conditions, occurring at a relatively lower value of the socioeconomic driver when dispersal conditions are high, as would be the case with a large proportion of the farms being sun farms, or many open habitats in the general region (e.g., open pastures). Furthermore, the qualitative nature of the trajectories suggests a classical "critical slowing down" of the system near the tipping point, not a surprising behavior for this sort of model (Sheffer et al., 2012), and offering the hope that prediction of an epizootic might be possible with a regional monitoring program. On the contrary, the classical critical slowing down does not accompany the other bifurcation (the switch from figure 2b to figure 2c), meaning that available leading indicators of the potential regime shift from "possible control" (fig 2b) to "no possibility of control" (fig 2c) may be difficult to find.

The sort of bifurcation diagram presented in figure 3 could be repeated for each of the seven parameters in the system. Not surprisingly, given the basic dynamic behavior of the system (see figure 2), the same threshold-like behavior emerges whatever parameter is used as the tuning parameter (the abscissa in figure 3). The behavior can be summarized by noting that increases in $n$, $M$, $a$ and $m$ and decreases in $p$, $E$ and $e$, result in the eventual emergence of a tipping point and at least the potential of an epizootic. Furthermore, again obvious from an examination of figure 2, the initiation point of the disease will partly determine the position of the tipping point.

Since coffee is reported to be one of the most traded commodities in the world, (Pendergast, 1999) and the base of economic support for millions of small farmers (Bacon, 2004), the 2012/2013 epizootic of the coffee rust disease extending from Mexico to Perú has justifiably attracted considerable attention (Cressey, 2013). This case has been especially severe and highlights the unpredictability of the disease – it has, since its arrival in the 1980s never been this severe. Some reports anticipate as much as a 40% - 50% reduction in yield over the region, potentially affecting many coffee-producing nations in Latin America (Inside Costa Rica, 2013). For example, our regular monitoring of a small plot in southern Mexico reflects the severity of the current epidemic, where over 60% of the plants experienced more than 80% defoliation and almost 9% expired completely. In late 2013 the Costa Rican government signed an emergency decree allocating $4 million to fight the spread of the disease (Inside Costa Rica, 2013). Similar patterns are reported anecdotally from Mexico to Peru (Cressey, 2013).

Dealing with the problem is challenging, with current recommendations focused on fungicides and phytosanitation procedures based on uncertain background information. Indeed, the complexity of this disease has been challenging for conventional disease control strategies that are largely based on a therapeutic emergency response paradigm (Kushalappa and Eskes, 1989; Avelino et al., 2004; Soto-Pinto et al., 2002). The model presented here suggests that it could be the larger ecological structure of the agroecosystem, both local and regional aspects, that needs to be considered, echoing the many recent calls for a more nuanced approach to the management of ecosystem services in general (Vandermeer et al., 2010), and pest control specifically (Lewis et al., 1997).